\documentclass{acm_proc_article-sp}
\usepackage{amsmath,amstext,verbatim,epsfig}
\usepackage{graphicx,psboxit}
\usepackage{epstopdf}
\usepackage{psfrag}

\begin{document}
\title{Publication Induced Research Analysis (PIRA) - Experiments on Real Data}

\numberofauthors{4}
\author{
\alignauthor
Gerard Burnside\\
       \affaddr{\normalsize Alcatel-Lucent Bell Labs Research}\\
       \affaddr{\small Route de Villejust}\\
       \affaddr{\small 91620 Nozay, France}\\
       \email{\small gerard.burnside@alcatel-lucent.com}
\alignauthor
Dohy Hong\\
       \affaddr{\normalsize Alcatel-Lucent Bell Labs Research}\\
       \affaddr{\small Route de Villejust}\\
       \affaddr{\small 91620 Nozay, France}\\
       \email{\small dohy.hong@alcatel-lucent.com}
\alignauthor
Son Nguyen-Kim\\
       \affaddr{\normalsize Alcatel-Lucent Bell Labs Research}\\
       \affaddr{\small Route de Villejust}\\
       \affaddr{\small 91620 Nozay, France}
       \email{\small nguyenkims@gmail.com}
\and
\alignauthor
Liang Liu\\
       \affaddr{\normalsize Alcatel-Lucent Bell Labs Research}\\
       \affaddr{\small Route de Villejust}\\
       \affaddr{\small 91620 Nozay, France}
       \email{\small mmhancxt@gmail.com}
}

\date{}

\maketitle
\begin{abstract}
This paper describes the first results obtained by implementing
a novel approach to rank vertices in a heterogeneous graph, 
based on the PageRank family of algorithms and applied here
to the bipartite graph of papers and authors as a first evaluation
of its relevance on real data samples. \newline
With this approach to evaluate research activities,
 the ranking of a paper/author depends
on that of the papers/authors citing it/him or her. \newline
We compare the results against existing ranking methods (including
methods which simply apply PageRank to the graph of papers or the graph
of authors) through the analysis of simple scenarios based on a real
dataset built from DBLP and CiteseerX. \newline
The results show that in all examined cases the obtained result is most
pertinent with our method which allows to orient our future work to
optimizing the execution of this algorithm.


\end{abstract}
\category{G.2.2}{Discrete Mathematics}{Graph Theory}[Graph algorithms]
\category{F.2.2}{Analysis of algorithms and problem complexity}{Nonnumerical Algorithms and Problems}[Sorting and searching]
\category{H.3.3}{Information storage and retrieval}{Information Search and Retrieval}[relevance feedback, search process]

\terms{Algorithms}
\keywords{publication, citation, ranking, graph, dataset.}
\begin{psfrags}
\section{Introduction}
This paper explores the implementation of existing work on a parameterized random
journey of a bipartite paper-author graph, which was defined in \cite{hong}. 
We recall that the exploration of a partial graph derived from this global one
was considered in past research in many papers
(for instance, the author graph in e.g. \cite{rad, zyc, liu, EY, fiala, YD}, or 
the paper graph \cite{chen, maslov, gori}, or a partial
joint graph in \cite{ma, zhou, yan, zhang}).

The main focus here is to revisit the analysis of different approaches in a more
systematical way than that considered in \cite{hong} and connect those results to the
real situations that were observed in the dataset that we built specifically for this study.

Building a large coherent dataset for that purpose was not straightforward:
the main difficulty was to obtain the information on the reference (or citation) list of a paper,
even if there are many public information such as Google Scholar, because of the legal issue
related to web site crawling.

In Section \ref{sec:model}, we recall briefly the algorithm that is
studied and will be compared to other metohds. In Section \ref{sec:data}, we describe how
the dataset was built for this evaluation.
Finally, Section \ref{sec:eval} illustrates and compares different approaches by outlining
major tendencies but also by pointing out several specific cases where the ranking differs
between our algorithm and another method.

\section{Model}\label{sec:model}
In this paper, we revisit the ranking algorithm proposed in \cite{hong}:
as in \cite{hong}, we give here a description of the algorithm based on the
random walk point of view for a better intuition and clarity. We recall that
the algorithm definition and the way the algorithm is solved are two
different problems. In this paper, the random walk approach is used to
obtain the score of each node (paper or author): as for the classical PageRank problem, we
could have used any other approach such as a power iteration, but such a consideration
is out of the scope of this paper and will be studied in a subsequent paper.

\subsection{PIRA algorithm}
The main ideas of \cite{hong} is to use the PageRank extension in the bipartite
graph of author and paper nodes where the notion of what we call here
{\it p-weight} and {\it c-weight} has been implicitly used. 

\subsubsection{Probability weight}\label{subsec:edgeProb}
{Probability weight} ({p-weight}) is a property associated 
to an edge for deciding the probability that this edge is chosen when the \textit{surfer} 
arrives at its origin node. 
For example, in Figure \ref{fig:edgeProb}, when the 
surfer arrives at the node A, he has a probability $2/3$ to go to C, and 
$1/3$ to go to B. The value of the property \textit{p-weight} is therefore {relative}, 
i.e. it is only useful when placed with other edges: if a vertex $ v $  has only one outgoing edge, 
then the probability of choosing this edge when a visitor arrives at $ v $ is always $1$
(conditional to the damping probability). 

\begin{figure}[h]
        \centering
        \includegraphics[width=6cm]{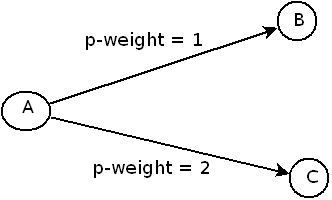}
        \caption{Edge probability weight}
        \label{fig:edgeProb}
\end{figure}
    
In the author paper graph, this p-weight of an edge from an author $a$ to a paper $p$ may be
represented as proportional to the time he spent to write $p$: 
if $p$ is written by three authors, then this value is $\frac{1}{3}$. More generally:

    \begin{equation*}
        pWeight(e) = \frac{1}{nbAuthors(p)}
    \end{equation*}

where $e$ is an edge linking an author to one of his paper $p$, and nbAuthors($p$) 
is the number of co-authors of $p$.

\subsubsection{Counter weight}  \label{subsec:counterProb}  
In the original PageRank, when arriving at a vertex $v$, we increment the counter
$counter(v)$ by one. 
A property that we call {counter weight} ({c-weight}) can be associated to an 
edge which decides the quantity of increment when the \textit{surfer} arrives at its end
using this edge:
    \begin{equation*}
        counter(v) = counter(v) +  cWeight(e) 
    \end{equation*}
where $e$ is an edge pointing to $v$.
    
The c-weight of an edge $e$ from a vertex $A$ to a vertex $B$ represents the \textit{weight} 
that $A$ \textit{gives} to $B$. For instance, if we want that a paper does not receive any score 
from its author (which can be considered as a self-evaluation), 
the c-weight of an edge linking an author to a paper may be set to 0:
    \begin{equation*}
        cWeight(e) = 0
    \end{equation*}
where $e$ is an edge linking an author to one of his papers.

Because the counter weight effect is not propagated through the links, this can be
seen as a reweighted score of the eigenvector (limit) obtained in the case where 
all c-weights are equal to one.

\subsection{Pseudo code}

The pseudo code below follows closely the flow execution of the algorithm proposed in \cite{hong}
as PR-G.

The \textit{restarting\_weight} is added to $counter(v)$ where $v$ is the vertex 
from which the visitor starts/restarts the random walk. The \textit{cite\_weight}, 
\textit{wrote\_weight}, \textit{isWrittenBy\_weight} are the c-weights associated to the
\textit{cite}, \textit{wrote} and \textit{isWrittenBy} edges respectively. 
The methods \textbf{a2p}, \textbf{p2p} and \textbf{p2a} stand for the jump from author to paper, 
paper to paper and paper to author respectively. 
In \textit{a2p}, the visitor picks a paper by taking into account the p-weight defined in 
Section \ref{subsec:edgeProb}.\newline
\textit{df} stands for the damping factor ($df = 1- d$ is the probability that the reinitialization
is triggered), and \textit{theta} is the probability that the \textit{surfer} 
follows a citation link when he arrives at a paper (otherwise an isWrittenBy link is chosen). 
\begin{verbatim}
init_all()
    choose a type: author or paper
    if (author) choose randomly an author a
                a2p(a , restarting_weight)
    else  choose randomly a paper p
          p2p(p, restarting_weight)

a2p(author,  weight)
    add weight to author 
    if (df)  init_all()		
    else  pick a paper p of a
          if (p exists)  p2p(p, wrote_weight)
          else init_all()

p2p(paper, weight)
    add weight to paper
    if (df)  init_all()		
    else  pick a cited paper p'
          if (p' exists)
                if (theta) p2p(p', cite_weight)
                else p2a(p, cite_weight)
          else init_all()

p2a(paper, weight)
    add weight to paper
    if (df)  init_all()
    else  pick randomly an author a of p
          if (a exists)
              a2p(a, isWrittenBy_weight)
          else init_all()

main()
    init_all()
\end{verbatim}

\section{Dataset}\label{sec:data}

\subsection{Construction phase}
In order to validate/evaluate our approach, we built a dataset of the
author-paper graph from DBLP \cite{dblp} and CiteSeerX \cite{citeseerx,giles} as follow:
\begin{itemize}
\item{[Starting point]} we parsed an XML file of DBLP containing 980680 papers and 679282 authors
  to create an initial author and paper sets (at this point the graph contains only \textit{wrote} edges);
\item{[Crawling links]} from the 980680 papers of DBLP, we crawled the CiteSeerX website to get information
  on the citation list of each paper (\textit{cited by}): from this we got only about 7\% successful answers (the rest is not
  found) amounting to 67772 papers;
\item{[References outside DBLP]} the dataset has been completed with papers (and authors) that cite the 67772 papers.
\end{itemize}

\begin{figure}[h]
        \centering
        \includegraphics[width=\linewidth]{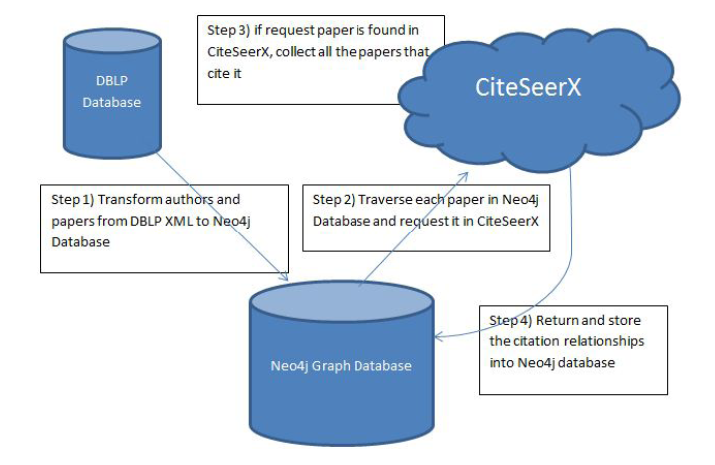}
        \caption{Dataset building process}
        \label{fig:build}
\end{figure}

The result of the above process is a dataset of 246039 authors (73241 from DBLP)
and 281207 papers (67772 from DBLP). In the following, while running the ranking algorithm on the
$246039 + 281207 = 527246$ nodes, the ranking results are only shown for those in DBLP because 
the authors and papers outside DBLP have no incoming citation links.
We could have iteratively crawled to get the citation list of the papers outside DBLP, but this
would have required an exponentially increasing data collection time and we estimated that 
the amount of data already collected was sufficient to illustrate several real-life scenarios.

\subsection{Dataset statistics}

This subsection provides some statistics on the dataset that was constructed as described above.

The average number of publications for authors from DBLP is 6.95 whereas it is only 1.7 for authors outside of DBLP.
\begin{figure}[!h]
        \centering
        \includegraphics[width=\linewidth]{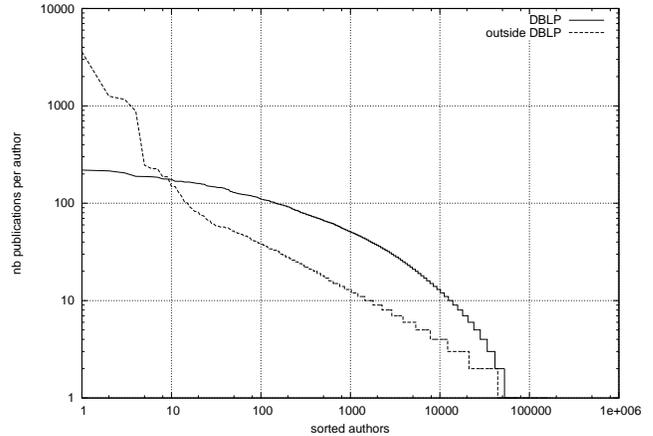}
        \caption{Number of publications per author}
        \label{fig:nbPub}
\end{figure}
Figure \ref{fig:nbPub} shows a logscaled comparison of the repartition of the number of 
publications per author for authors inside and outside of DBLP.
Note that the first 9 authors outside DBLP with over hundreds of publications are actually
parsing errors from the CiteSeerX dataset (such as: et al., Ph D, Student Member, etc.). Because we
do not rank those authors we preferred at first to leave them since they represent a known lack of information
in our dataset (which we may decide to correct at a later time).

The average number of co-authors per paper is similar for both DBLP and non DBLP papers ($\approx$2.85).
\begin{figure}[!h]
        \centering
        \includegraphics[width=\linewidth]{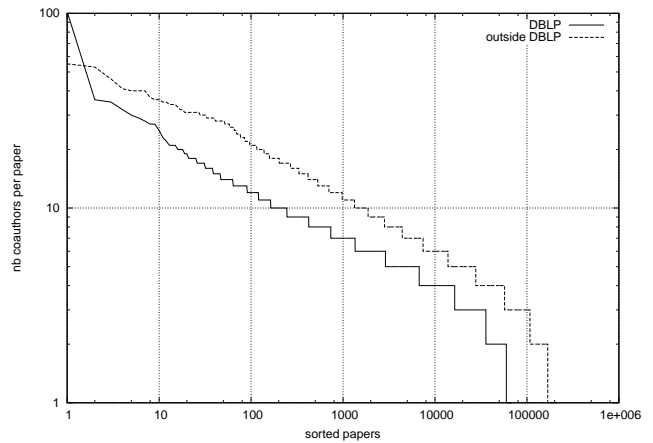}
        \caption{Number of coauthors per paper}
        \label{fig:nbCoAuth}
\end{figure}
Figure \ref{fig:nbCoAuth} shows a logscaled comparison of the repartition of the number of co-authors per paper inside and outside of DBLP. 
Note that the paper from DBLP with 102 authors is in fact some kind of compilation of many scientists' work.

The total number of citation links in our dataset is 631113, of which 121688 are citations 
between papers both in DBLP (the rest being citations from outside DBLP to a paper in DBLP: 
by construction we cannot have a citation towards a paper outside DBLP).
\begin{figure}[!h]
        \centering
        \includegraphics[width=\linewidth]{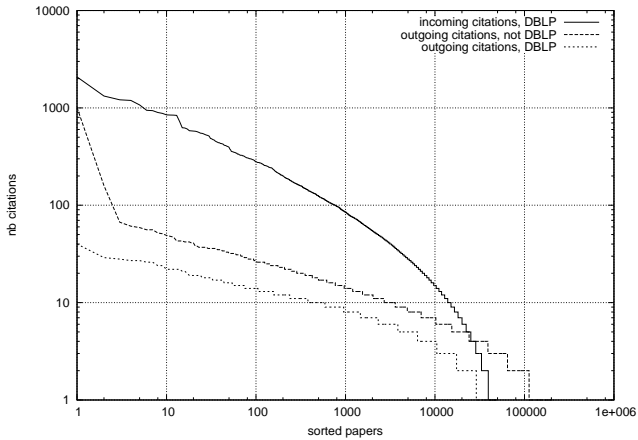}
        \caption{Number of citations per paper}
        \label{fig:nbCit}
\end{figure}
Figure \ref{fig:nbCit} shows the number of outgoing citations from papers inside and outside DBLP (with no surprise the non-DBLP line stays above the DBLP one); we also included in the figure the incoming citations line (which makes sense only for DBLP papers by construction).

\subsection{Data validation}
The process of building a clean dataset is in fact not obvious: in particular, we encountered
the very well known problem of the author name disambiguation. This problem may be partially solved
by assuming that an author is likely to cite his own papers and based on the co-authorship information:
for instance, if a paper A of an author J. YYY cites a paper B of John YYY, we may assume that
the author of paper A is John YYY. Also, when J. YYY and John YYY have a common co-author,
we may also assume that J. YYY is John YYY for the co-authored papers. Unfortunately for other papers which did not
meet one of the above conditions we cannot assume that J. YYY is John YYY, therefore our dataset
still contained many ambiguous authors (with initials) which we preferred to leave as it is in order not
to introduce false authorships (i.e. when in doubt we preferred to have duplicate authors rather than
merging two actual distinct authors into one).

As pointed out, such disambiguation methods are far from being enough, but such a consideration is out of the
scope of this paper.

\section{Evaluation}\label{sec:eval}

\subsection{Notation}

In order to compare the proposed solution to existing measures, 
several theoretical scenarios are considered. 
Those scenarios are also found in our database, which shows that those scenarios
are not only fictive ones. The following notation will be used:
\begin{itemize}
    \item \textbf{Pub.} of an author: the number of publications of the author;
    \item\textbf{Cit.} of an author: the number of times his papers have been referenced;
    \item\textbf{Hind} of an author: the H-index;
    \item \textbf{PR-A} of an author: the PageRank score of this author on the \textit{author graph};
    \item \textbf{PR-P} of an author: the PageRank score of this author resulting from the \textit{paper graph} 
               (i.e. each author receives the fraction of the PageRank score of his/her papers corresponding 
                to an even distribution among each paper's authors);
    \item \textbf{PIRA} of an author: the score of this author in the PR-G algorithm.
\end{itemize}
Note that there are in fact many possible ways to implement the PageRank variants on
the author graph. In this paper, we considered the one derived from PIRA when constraining
the random walk on the paper graph following the citation link exactly one time.

\subsection{Preliminary analysis}


In this section we will comment some general impact of the ranking differences due to the
limited dataset information, which may be seen as a boundary impact.

\subsubsection{The weight of external papers}
In our dataset, about 70-75\% papers/authors are {\em external} (i.e. outside \textsc{dblp}), therefore their impact
on the ranking score is dominant.
This may result in limited differences when considering variants of ranking algorithm,
in particular because the random walk path is not {\em deep} enough.

\subsubsection{Missing data on the referenced papers list} 
Because we had only a subset of the scientific literature and also because of the way 
we constructed our dataset, the average number of references to another paper in our dataset 
is only slightly over 2 (see also Figure \ref{fig:nbCit}), whereas the actual number of references per paper is usually in the 
10-30 range (only 4188 papers in our dataset have more than 10 references to other 
papers in our database, they represent less than two percent of the total number of papers).

We suspected that having only a single or two outgoing citation links was going to introduce 
bias in some cases and this is why we implemented a \textsc{minimum\_citation\_count} variable 
to simulate a minimum number of outgoing references and therefore dilute the chances to pick 
a specific paper (the only one available) when following the citation link 
(when a "fake" paper is chosen, we restart the random walk from any paper in the dataset).

We have run our PIRA algorithm with the variable \textsc{minimum\_citation\_count} set to 0 
and then to 10 and compared the results.
As we suspected, about 20 percent of papers ranked in the top one percent with the variable 
set to zero were ranked out of the top one percent when this variable is introduced and set to 10.

We then investigated a few examples of papers which were ranked significantly lower when 
setting the variable \textsc{minimum\_citation\_count} to 10.

For example author 15510 ({\em Gilles Brassard}) was ranked 159 without \textsc{minimum\_citation\_count} 
and went down to rank 960 when imposing a minimum of 10 references per paper.

\begin{figure}[h!]
\centering
\includegraphics[width=\linewidth]{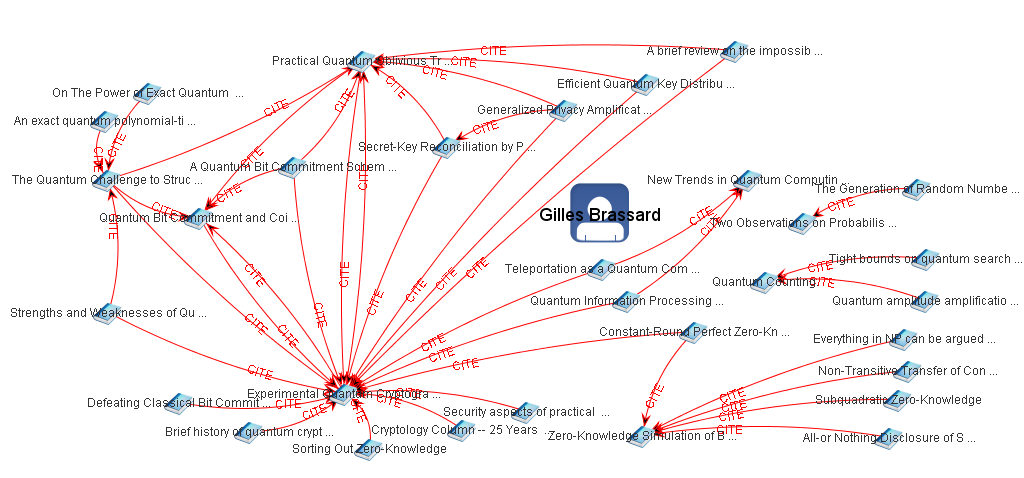}
\caption{Author 15510 and self-citation}
\label{fig:sample15510}
\end{figure}

Figure \ref{fig:sample15510} shows all the papers written by Gilles and their citation links among themselves.
Two things stand out in the figure: Gilles wrote a decent amount of papers 
(his publication score is 41) and most of his papers cite a previous paper of his 
(only nine papers did not cite other papers - they are not displayed in this figure).
When not imposing a minimum amount of references it is obvious that each time we end 
up on one of his paper and choose to follow the citation link we will have no choice 
but to jump to another one of his paper, thereby mechanically increasing its PIRA  score.
But this is not quite enough to explain the drastic difference in ranking: a closer 
inspection into which papers most influenced Gilles' score shows that one paper 15508 
({\em The quantum challenge to Structural Complexity}) had the most influence on his PIRA score 
and paper 15508 itself inherited most of its score from paper 86162 
({\em Quantum complexity theory} - which cites only two papers, the other being also 
written by Gilles) which in turn inherits it from paper 86399 
({\em A fast quantum Mechanical Algorithm for Database Search} - which cites only 86162), 
the latter two papers having a PIRA ranking of respectively 42 and 41.
This chain of citations is heavily diluted when introducing 
a \textsc{minimum\_citation\_count} of 10 (because two citation jumps will decrease the 
probability by two orders of magnitude) and this explains most of the ranking difference.

Another example which corroborates the fact that an indirect citation chain 
is the major factor for the down-ranking caused by \textsc{minimum\_citation\_count} is author 30803 
(Takuo Watanabe, whose rank goes from 177th down to 2030th) which has a much more modest publication 
score of 11 and which also inherits most of its score from a single paper 30802 
({\em Hybrid Group Reflective Architecture for OO Concurrent Reflective Programming}) 
which is cited by two well ranked papers:
\begin{itemize}
  \item 52911 ({\em Aspect-Oriented programming}) with a PIRA rank of 28, and which only cites 3 papers from our dataset;
  \item 30772 ({\em An Overview of AspectJ}) with a PIRA rank of 32  and which also cites only 3 papers.
\end{itemize}
Note that when comparing PIRA with PR-P (cf. section \ref{subsubsec:prp}), we have set
\textsc{minimum\_citation\_count} to 0 because we did not have the equivalent parameter on the 
existing PageRank implementation that we used.

\subsection{Global comparison}
In this section, we address a global comparison of ranking methods.
Figure \ref{fig:P-Score} shows the difference in percentage of the X best ranked authors
w.r.t. the number of publications. The ranking by the number of publications is probably
the measure which differs the most to all others: it is likely to be directly
proportional to the quantity of effort spent by an author. We can observe that the differences
are more visible with PR-A, Cit and PIRA which are intuitively the most qualitative ones.
We see that PR-P is the closest (below top 15\%): this can be explained by the fact that with
PR-P each published paper has a minimum score, therefore the score of an author is above
a linear function (before normalization) w.r.t. the number of publications.

\begin{figure}[h]
    \centering      \includegraphics[width=\linewidth]{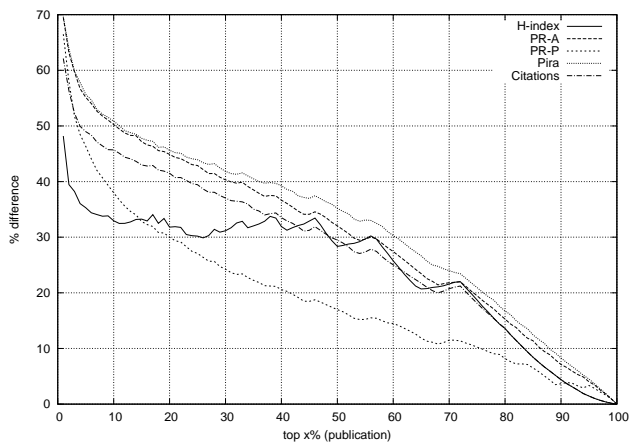}
    \caption{Comparison of x best ranked w.r.t. Pub.}
    \label{fig:P-Score}
\end{figure} 

Figure \ref{fig:C-Score} shows the same comparison but this time w.r.t. the number of citations: we can see that
PR-A is the closest to citations based ranking: the explanation is that PR-A explores
the paper citation links only once between authors who cite without the score inheritance between
authors distant by more than one citation relationships: as a consequence, the ranking is close to the
local counter of the number of citations.

\begin{figure}[h]
    \centering      \includegraphics[width=\linewidth]{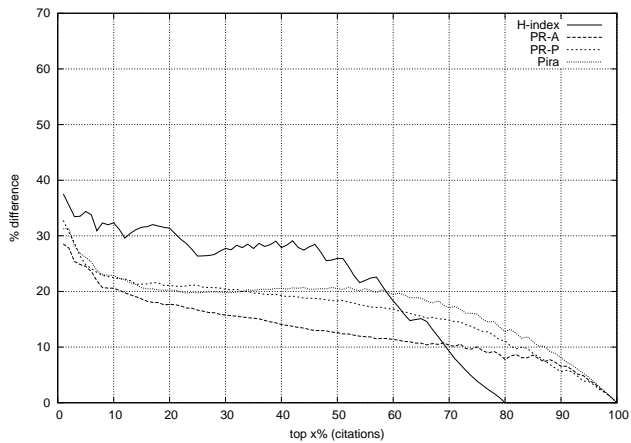}
    \caption{Comparison of x best ranked w.r.t. Cit.}
    \label{fig:C-Score}
\end{figure} 

The fact that in both cases (Figure \ref{fig:P-Score} and Figure \ref{fig:C-Score}) the best ranked
authors are less differentiated between different approaches is mostly due to the graph property 
(Zipf/power law type links distribution, cf. Figure \ref{fig:nbCit}) 
of the dataset: there are authors who will be in the top
1\%, 2\% etc whatever the measure (authors having a lot of publications and citations).
Note also that by definition, the difference on the top 100\% is always equal to 0.

\subsection{Specific cases comparisons}
\subsubsection{Citations count vs PIRA ranking}\label{subsubsec:cit}

In order to evaluate the ranking differences among authors between citations count and PIRA, 
we constructed a file containing the first thousand DBLP authors ranked with citations count 
and compared their rank in PIRA. See Figure \ref{fig:sample83130}  where the x-axis 
is the citations count ranking and the y-axis is for each author the difference between its 
citations rank and its PIRA rank.\newline
What we did was then to identify points which would stray far from the mass and investigate the rank difference.\newline
We started by identifying author 70411 which had a citation rank of 167 and 
a PIRA rank of 1613.
A short investigation showed that most of its rank was inherited from paper 
83130 which has 570 citations in our dataset. Interestingly enough this paper was 
co-authored by eight authors which are all outlined in  Figure \ref{fig:sample83130} 
because of their ranking differences between citations count and PIRA, 
which can easily be explained by the fact that in PIRA the weight of this paper will be 
divided evenly between the eight authors whereas citations count will freely distribute 
all this paper's weight to all its authors (in fact it can be noted that some of these authors 
received citations only for this paper, and still ranked 316th and 317th with citations count!).
\begin{figure}[h]
\centering
\includegraphics[width=\linewidth]{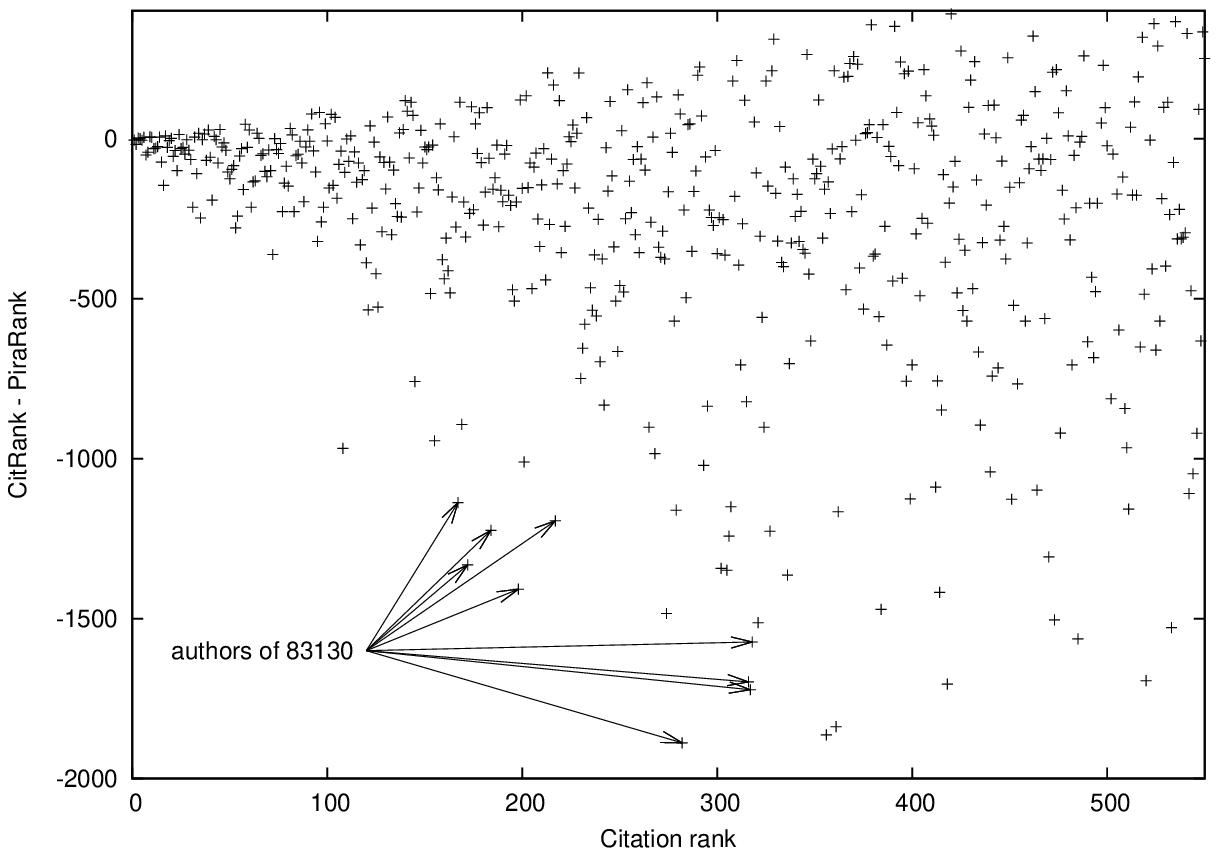}
\caption{Rank$_{citationCount}$ $-$ Rank$_\textsc{pira}$}
\label{fig:sample83130}
\end{figure}



Another point that stood out was author 11409 which is ranked 390 positions higher with PIRA than with citations count, this is caused by the fact 
that his main paper 74463 (of which he is the unique author, which helps rank him higher in PIRA 
compared to multiple authors of a paper with the same citation rank) was cited by a number of papers that were 
themselves quite cited (100+ times each), among which is 74457 written by author 34024 who 
is also well-ranked (both PIRA and citation) and which cites only two papers.

This illustrates clearly the impact of the recursive score inheritance of the PageRank approach which does not exist with 
local counters such as citations count, which brings us to the comparison of PIRA with PageRank applied on the papers graph (PR-P).



\subsubsection{PR-P vs PIRA ranking}\label{subsubsec:prp}

As in the previous section we have plotted the first thousand ranked papers with PR-P against
the rank difference with PIRA (see Figure \ref{fig:PrpPira}).
This helps to quickly identify papers whose rank is greatly modified by the algorithm change.
A first remark on Figure \ref{fig:PrpPira} is that most \textit{anomalies} correspond to a case where the
rank with PIRA is significantly worse than with PR-P.
\begin{figure}[h!]
    \centering      \includegraphics[width=\linewidth]{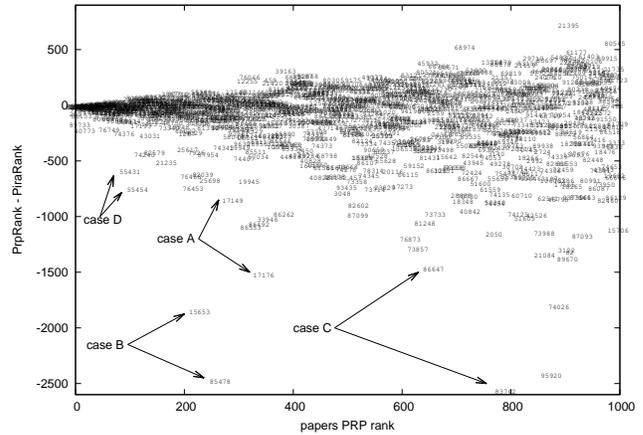}
    \caption{Ranking difference PRP-Pira}
    \label{fig:PrpPira}
\end{figure} 

When picking a paper among the \textit{anomalies} and using our specifically built graphical
tool (see Figure \ref{fig:useCaseA}) to investigate the reasons behind the ranking difference, we
have always found that at least one other paper was involved in the ranking difference and
this is why we grouped papers by pairs (cases A, B, C and D) in Figure \ref{fig:PrpPira}.

\begin{figure}[h!]
    \centering      \includegraphics[width=\linewidth]{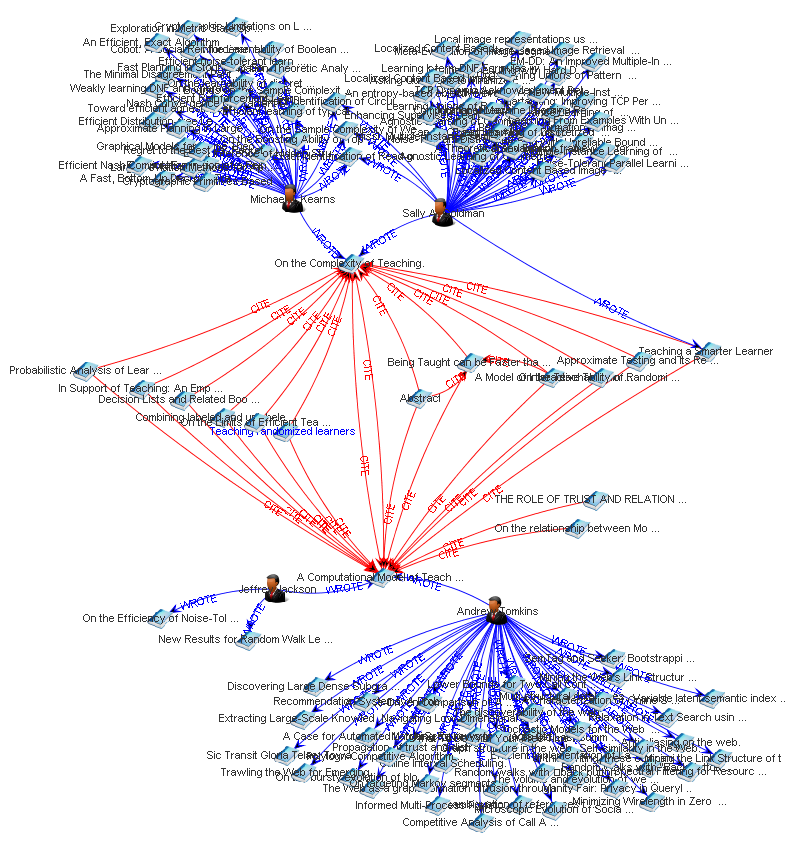}
    \caption{Use-case A in detail}
    \label{fig:useCaseA}
\end{figure} 

All the papers which strayed from the mass in Figure \ref{fig:PrpPira} have the particularity that
they cite a paper which cites themselves back. Although counter-intuitive (a paper should only be
able to cite papers from the \emph{past}), we have found many examples of this, confirmed
 by manually visiting CiteSeerX's website, usually because a revised paper \emph{X} will have
cited a paper \emph{Y} which cited an earlier version of \emph{X}.
Because most cases found in the \textit{anomalies} are similar (citation loop between two papers), we will
explain the difference using \emph{case A}. \newline
As can be seen in Figure \ref{fig:useCaseA}, the structure consists of two papers citing each other (and citing no other papers):
17176 (\textit{A Computational Model of Teaching}) and 17149 (\textit{On the Complexity of Teaching},
of which a preliminary version appeared in the Proceedings of \textit{4th Annual Workshop of Computational
Learning Theory}) and which are both cited by a relatively small (14) number of papers.
What happens intuitively with PageRank on the graph of papers is that once the \emph{random surfer} arrives on one of these 16
papers (the aforementioned fourteen plus our two PR-P inflated ones) it will then be trapped and jump only between paper
17176 and 17149 (until the damping factor takes effect), thereby artificially incrementing those two papers' scores.

What happens when we introduce the random walk on the \emph{bipartite graph} (\textit{authors+papers}) is that
we move from a paper to to an author with probabilty $1 - theta$ (usually 0.3 in our executions) and once on an
author the next jump will be uniformally chosen among its publications, and Figure \ref{fig:useCaseA} clearly
shows that 3 of the 4 authors have 30+ publications each, thereby removing the \textit{trap} that existed with PR-P.


\subsection{Generic comparison}\label{sec:compa}

\subsubsection{Paper quality}
The number of publications measure does not take into account 
the {paper quality} (the notion of \textit{quality} is approximate, 
since we first need to define a pertinent ranking for papers, 
then a quality paper will be a highly ranked one. For now, a paper is considered of quality 
if it is cited by an important number of papers): an author who has written a referential 
paper is less ranked than another who wrote two unknown papers 
(this can happen with old time, classic authors who published in general  less than the actual ones. 
This could be due to the lack of information we have  about ancient journals or conferences). 

\begin{figure}[h!]
    \centering      \includegraphics[width=\linewidth]{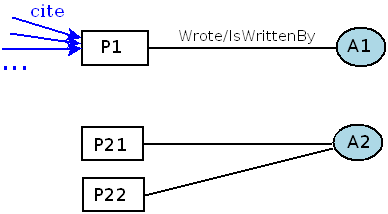}
    \caption{Paper quality. With Pub., A1 is less ranked than A2.}
    \label{fig:qualityPaper}
\end{figure} 

In our database, we have found (see Table \ref{tab:paperQuality}) two authors who illustrate the above scenario: one publishes a lot but receives nearly zero citation while the other wrote only one high quality paper.
\begin{table}[htbp]
    \caption{Quality of paper}
    \begin{center}
        \begin{tabular}{|c|c|c|c|c|c|}
            \hline
            ID & Name & Pub & Cit & PRA & PIRA \\ \hline
            20 & Dorothy E. & 1 & 313 & 63.748 & 59.942 \\
               & Denning    &   &     &        &  \\ \hline
            40152 & Pedro  & 21 & 1 & 0.077 & 0.331 \\
             & Cabalar &  &  &  &  \\ \hline
        \end{tabular}
    \end{center}
    \label{tab:paperQuality}
\end{table}

\subsubsection{Number of co-authors}
With the publication and citation measures, the {number of co-authors} (of the measured author) is not taken into account. 
A person who writes single-handedly his paper is considered equal to one who published a paper with 10 other persons.
\begin{figure}[h]
\centering      
\includegraphics[width=\linewidth]{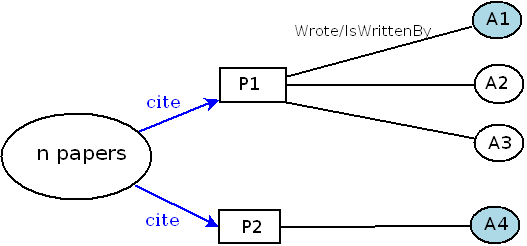}
\caption{Number of co-authors. With publication and citation measure, A1 and A4 are equally ranked.}
\label{fig:numberAuthor}
\end{figure} 
\subsubsection{Quality of citing papers}
With the publication, citation and PR-A measures, the {quality of citing papers} are not taken into account: 
a citation from a famous paper is considered the same as a citation from an unknown paper.
\begin{figure}[h]
    \centering      \includegraphics[width=\linewidth]{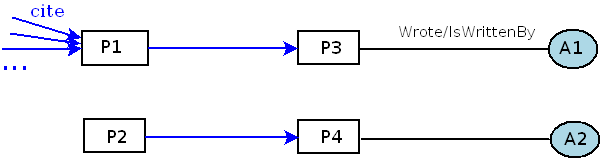}
    \caption{Quality of citing papers. With citation and publication measure, A1 and A2 are equally ranked.}
    \label{fig:qualityCitingPaper}
\end{figure} 

In our database, we have also found two authors who represents the above scenario: both publish 
just one paper which receives each just one citation, but the quality of the citing paper is very different: one receives 16 citations, the other has just one citation. 
As we can see on Table \ref{tab:citingPaperQuality}, only PIRA produces a good ranking.
\begin{table}[htbp]
    \caption{Quality of citing papers}
    \begin{center}
        \begin{tabular}{|c|c|c|c|c|c|c|}
            \hline
            ID & Name & Pub & Cit & PR-A & PIRA &distant  \\
	&	&	&	&	& 	&citations \\ \hline
            \tiny{85366} & \small{Soe} & 1 & 1 & 0.338 & 0.816 & 16 \\
	&\small{Myat Swe}&	&	&	& 	& \\ \hline
            \tiny{103666} & \small{Carlo}& 1 & 1 & 0.41 & 0.443 & 1 \\
	&\small{Jelmini}&	&	&	& 	& \\ \hline
        \end{tabular}
    \end{center}
    \label{tab:citingPaperQuality}
\end{table}
\subsubsection{Effect of self-citation}
In the publication and citation measures, the {effect of self-citation} is not taken into account. 
By writing a lot of papers, each one referencing the previous ones, an author can have a rather high ranking.
\begin{figure}[h]
    \centering      \includegraphics[width=\linewidth]{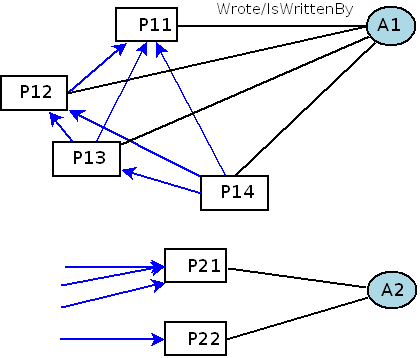}
    \caption{Effect of self-citation. With citation measure, A1 is better ranked than A2.}
    \label{fig:selfCitation}
\end{figure} 

In our database, we have also found two authors who represent the above scenario: 
one publishes several papers and receives only citations from papers written by himself, the other author publishes less, 
receives three times less citations but all of them come from exterior papers (which reveals a more legitimate impact).
\begin{table}[htbp]
    \caption{Effect of self-citation}
    \begin{center}
        \begin{tabular}{|c|c|c|c|c|c|c|}
            \hline
            ID & Name & Pub & Cit & PR-A & PIRA & cit. ext. \\ \hline
            \tiny{99898} &\small{Nicolai} & 10 & 9 & 0.669 & 0.526 & 0 \\ 
             &  \small{Czink} &  &  &  &  &  \\ \hline
            \tiny{77239} & \small{Christophe} & 1 & 3 & 2.008 & 1.038 & 3 \\ 
             &  \small{Lizzi} &  &  &  &  &  \\ \hline
        \end{tabular}
    \end{center}
    \label{tab:coAuthor}
\end{table}

\subsubsection{Summary}
The summary of what features a measure does or does not take into account is proposed in Table \ref{tab:author}. 
\begin{table}[htbp]
    \caption{Comparison table for different metrics}
    \begin{tabular}{|c|c|c|c|c|}
        \hline
        Criteria/Measures & \textbf{Pub} & \textbf{Cit} & \textbf{PR-A} & \textbf{PIRA} \\ \hline
        \textit{Paper quality} & No & \textbf{Yes} & Yes & Yes \\ \hline
        \textit{Number of co-authors } & No & No & Yes & Yes \\ \hline
        \textit{Citing papers' quality} & No & No & \textbf{No} & Yes \\ \hline
        \textit{Self-citation's effect} & No & No & Yes & {Yes} \\ \hline
    \end{tabular}
    \label{tab:author}
\end{table}

\section{Conclusion}\label{sec:conclusion}
In this paper, we revisited a global bipartite graph based ranking algorithm
for jointly ranking papers and authors and 
compared this ranking mechanism to existing metrics through simple
and generic cases to illustrate the improvements brought by this type
of ranking: a real dataset was built from DBLP and CiteseerX to illustrate
the results and to show how this global bipartite graph approach has 
the advantage of being qualitatively more relevant. \newline
Future work will focus on improving the algorithm in terms of performance
to obtain a significantly faster ranking compared to existing methods.


\end{psfrags}
\bibliographystyle{abbrv}
\bibliography{sigproc}

\end{document}